\documentclass[aps,pra,twocolumn,showpacs,superscriptaddress,amssymb,amsmath,amsmath, nobibnotes]{revtex4-1}
\usepackage{graphicx}
\usepackage{epstopdf}
\usepackage{bm}
\usepackage{hyperref}
\usepackage{comment}
\usepackage{multirow}
\usepackage{float}
\usepackage{times}
\usepackage{cancel}

\hypersetup{%
   pdfpagemode=None, 
   pdfstartpage=1,
   pdfmenubar=true,
   pdftoolbar=true,
   colorlinks = true,
   linkcolor=blue,
   citecolor=blue,
   urlcolor=blue,
   bookmarksopen=false
 }

\begin{document}
\title{Van der Waals molecules consisting of a zinc or cadmium atom interacting with an alkali-metal \\ or alkaline-earth-metal atom}

\author{Klaudia Zaremba-Kopczyk}
\email{klaudia.zaremba-kopczyk@fuw.edu.pl}
\affiliation{Faculty of Physics, University of Warsaw, Pasteura 5, 02-093 Warsaw, Poland}
\author{Micha{\l} Tomza}
\email{michal.tomza@fuw.edu.pl}
\affiliation{Faculty of Physics, University of Warsaw, Pasteura 5, 02-093 Warsaw, Poland}
\date{\today}

\begin{abstract}

Alkaline-earth-like transition-metal atoms such as Zn and Cd are promising candidates for precision measurements and quantum many-body physics experiments. Here, we theoretically investigate the properties of diatomic molecules containing these closed-shell atoms. We calculate potential-energy curves, permanent electric dipole moments, and spectroscopic constants for molecules consisting of either a Zn or Cd atom interacting with an alkali-metal (Li, Na, K, Rb, Cs, Fr) or alkaline-earth-metal (Be, Mg, Ca, Sr, Ba, Ra) atom. We use the \textit{ab initio} electronic-structure coupled-cluster method with single, double, and triple excitations combined with large Gaussian basis sets and small-core relativistic energy-consistent pseudopotentials for heavier atoms. We predict that the studied molecules in the ground electronic state are chemically reactive weakly bound van der Waals complexes with small permanent electric dipole moments. The present results may be useful for spectroscopy and application of the studied molecules in modern ultracold physics and chemistry experiments.

\end{abstract}

\maketitle

\section{Introduction}

Ultracold molecules have emerged in recent years as a versatile platform for studies of complex quantum phenomena~\cite{CarrNJP09,QuemenerCR12,BohnScience17}. The rich internal molecular structure and intermolecular interactions have been employed in studies of quantum many-body physics, allowing for the realization of many-body Hamiltionians of yet unexplored complexity~\cite{BaierScience16,GrossScience17}. The controllability of molecular collisions with external magnetic or electric fields, along with precise control over molecular quantum states, have enabled research on ultracold controlled chemical reactions~\cite{OspelkausScience10,TomzaPRL15,HuScience19}. Furthermore, the complexity of molecular structure provides novel possibilities for precision tests of fundamental physics, which include tests of fundamental symmetries, searches for spatiotemporal variations of fundamental constants, tests of quantum electrodynamics, and tests of general relativity, among others~\cite{DeMilleScience17,SafronovaRMP18}.

Ultracold molecules containing alkaline-earth-type atoms are promising candidates for high-precision measurement experiments~\cite{McDonaldNature16} and emerging quantum technologies~\cite{KondovNP19}, while alkaline-earth-type atoms have already served as important building blocks of high-precision physics~\cite{DereviankoRMP11}. For example, optical lattice clocks based on the $^1S_0 \rightarrow{}^3P_0$ transition in alkaline-earth-type atoms have played a substantial role in establishing current time and frequency standards~\cite{KatoriPRL03,LudlowScience08,LemkePRL09,LeTargatNatCom13}. Here, the main focus has been put on optical lattice clocks based on strontium~\cite{KatoriPRL03,LudlowScience08,LeTargatNatCom13}, ytterbium~\cite{PoliPRA08, LemkePRL09}, and mercury~\cite{HachisuPRL08,McFerranPRL12,YamanakaPRL15} atoms; however, recent proposals have brought attention to two other suitable candidates, zinc and cadmium atoms~\cite{OvsiannikovPRA16,DzubaJPhysB19,YamaguchiPRA19,PorsevPRA20}. Optical lattice clocks based on group-IIB atoms, such as Zn, Cd, and Hg, have been shown to exhibit reduced susceptibility to the black body radiation (BBR) as compared to Sr- and Yb-based clocks~\cite{HachisuPRL08,McFerranPRL12,SafronovaPRA13,YamanakaPRL15,OvsiannikovPRA16,YamaguchiPRA19}. With BBR being the major factor limiting the accuracy of atomic clocks, Zn and Cd atoms serve as promising alternatives to the currently operational Sr and Yb clocks~\cite{DzubaJPhysB19,PorsevPRA20}. In addition, optical lattice clocks based on alkaline-earth-type atoms are excellent systems for quantum simulations of many-body physics~\cite{KolkowitzNature17,GobanNature18}. Finally, optical clock transitions in divalent atoms have been suggested as a tool to explore potential variations in the fine-structure constant~\cite{AngstmannPRA04, HachisuPRL08} or establishing constraints on the value of the electron's electric dipole moment (EDM)~\cite{HachisuPRL08}.

The use of ultracold molecules based on alkaline-earth-type atoms provides further enhancement of sensitivities to the variations of fundamental constants or EDM effects~\cite{SafronovaRMP18}. In this context, one potentially interesting class of molecules is heteronuclear molecules composed of a closed-shell alkaline-earth-like atom interacting with an open-shell atom, such as an alkali-metal atom~\cite{ZuchowskiPRL10} or a halide~\cite{KozlovPRL06}. Such molecules have been proposed to be useful for measuring the variations in the proton-to-electron mass ratio~\cite{KajitaPRA11} and suggested as appealing candidates for searches of the electron's EDM~\cite{KozlovPRL06,MeyerPRA09,PrasannaaPRL15,SunagaPRA18,VermaPRL20}. Moreover, homonuclear dimers of alkaline-earth-type atoms also show prospects for precise measurements of the proton-to-electron mass ratio~\cite{ZelevinskyPRL08,KotochigovaPRA09}, while heteronuclear $^2\Sigma$-symmetry molecules have been proposed as quantum simulators with prospects for creating topologically ordered states~\cite{MicheliNatPhys06}. $^2\Sigma$-state molecules can be formed from ultracold mixtures of closed-shell and open-shell atoms, following recent experimental advances in studies of Yb+Rb~\cite{NemitzPRA09}, Sr+Rb~\cite{BarbeNP18}, Yb+Li~\cite{GreenPRX20}, and Yb+Cs~\cite{WilsonPRA21} combinations. 

In this work, we propose the formation of ultracold heteronuclear molecules composed of a transition-metal zinc or cadmium atom interacting with an alkali-metal or alkaline-earth-metal atom. The electronic structure of homonuclear dimers of group-IIB atoms, such as Zn$_2$, Cd$_2$, and Hg$_2$, has been the subject of theoretical studies~\cite{PetersonTCA05,PahlTCA11,WeiJCP13,UrbanczykIRPC17}, while heteronuclear molecules composed of a Hg atom interacting with an alkali-metal atom have been investigated both theoretically~\cite{ThielJCP03,SunagaPRA19} and experimentally~\cite{WitkowskiOE17,WitkowskiPRA18,BorkowskiPRA17}. To the best of our knowledge, molecules composed of a Zn or Cd atom interacting with an alkali-metal or alkaline-earth-metal atom have not yet been investigated in the literature. Transition-metal zinc and cadmium atoms, compared to the alkaline-earth-metal atoms, possess a richer structure of excited electronic states due to the possibility of electron excitations from the $d$ subshell. While the ground-state electronic structure of such zinc- or cadmium-containing molecules resembles the electronic structure of alkaline-earth-metal or alkali-metal--alkaline-earth-metal molecules, the richer electronic structure of constituent atoms would have its reflection in a more complex structure of excited electronic states, which may find application in precision measurements~\cite{SafronovaRMP18}. Additionally, zinc or cadmium atom interacting with other atoms may form weakly bound van der Waals molecules that may potentially be used as precise probes of new gravity-like forces~\cite{SalumbidesPRD13,BorkowskiSciRep19}. The ongoing progress in laser cooling and trapping of cadmium atoms~\cite{XuPRA04,BrickmanPRA07,KanedaOL16,YamaguchiPRA19} further motivates our investigation.

Here, we theoretically investigate the ground-state properties of diatomic molecules composed of either a Zn or Cd atom interacting with an alkali-metal (Li, Na, K, Rb, Cs, Fr) or alkaline-earth-metal (Be, Mg, Ca, Sr, Ba, Ra) atom. We use state-of-the-art electronic structure methods to calculate the potential-energy curves (PECs) and spectroscopic constants for the investigated molecules. We predict that the considered molecules in the ground electronic state are weakly bound van der Waals complexes, which are chemically reactive. They possess rather small permanent electric dipole moments, despite Zn and Cd atoms having electronegativity significantly larger than that of alkali-metal and alkaline-earth-metal atoms. In this way, the present study extends the range of species available for ultracold molecular studies.

This paper is constructed as follows. Section~\ref{sec:methods} introduces the \textit{ab initio} electronic structure methods employed in our calculations. Section~\ref{sec:results} presents and analyzes the obtained numerical data, including the potential-energy curves and electric properties of the investigated molecules. Finally, Sec.~\ref{sec:summary} summarizes our paper.

\section{Computational details}
\label{sec:methods}

In order to calculate the potential-energy curves within the Born-Oppenheimer approximation, we employ the closed-shell and the spin-restricted open-shell coupled-cluster methods restricted to single, double, and noniterative triple excitations [CCSD(T)]. Next, we include the full iterative triple-excitation correction, $\Delta$T, calculated with the use of the coupled-cluster method restricted to single, double, and full triple excitations (CCSDT). We obtain the counterpoise-corrected interaction energies within the supermolecule approach~\cite{BoysMolPhys70}.

We use the small-core scalar-relativistic energy-consistent pseudopotentials from the Stuttgart/K\"oln library, ECP$n$MDF, to describe $n$ inner-shell electrons of studied transition-metal atoms, and heavier alkali-metal and alkaline-earth-metal atoms (ECP10MDF for Zn, K, and Ca; ECP28MDF for Cd, Rb, and Sr; ECP46MDF for Cs and Ba; and ECP78MDF for Fr and Ra)~\cite{FiggenCP05, LimJCP05,LimJCP06}. This approach treats only the electrons from the two outermost shells of a given atom explicitly [i.e.,~$3s^23p^63d^{10}4s^2$ from Zn, $4s^24p^64d^{10}5s^2$ from Cd, $(n-1)s^2(n-1)p^6{n}s^1$ from alkali-metal, and $(n-1)s^2(n-1)p^6{n}s^2$ from alkaline-earth-metal atoms], and hence, it allows us to use larger basis sets for more accurate molecular calculations. We correlate all remaining electrons. For the presented computations at the CCSD(T) level of theory, we employ the corresponding correlation-consistent polarized weighed core-valence quintuple-$\zeta$ quality basis sets (aug-cc-pwCV5Z-PP~\cite{PetersonTCA05,HillJCP17} with ECP and aug-cc-pwCV5Z~\cite{PrascherTCA11} for Li, Na, Be, and Mg) augmented by the set of the $[3s3p2d2f1g]$ bond functions. To account for the full triple-excitation correction ($\Delta$T), we perform electronic structure calculations at the CCSDT level of theory with the use of valence-only triple-$\zeta$ quality basis sets (aug-cc-pVTZ for Li, Na, Be, and Mg atoms, and aug-cc-pVTZ-PP for the remaining atoms).

Additionally, for two representative systems, an open-shell RbZn molecule and a closed-shell SrZn molecule, we carry out convergence tests to analyze the accuracy of the obtained interaction energies and confirm the optimal method and basis sets for the remaining calculations. To this end, we compute interaction energies using the CCSD(T) method and aug-cc-pwCV$n$Z-PP basis sets with $n =$ D, T, Q, 5. We use these basis sets to extrapolate the interaction energies to the complete-basis-set (CBS) limit and show that adding a bond function (BF) to aug-cc-pwCV5Z-PP basis sets allows us to reproduce the CBS limit accurately. Next, we obtain the full iterative triple-excitation correction ($\Delta$T), given as a difference between interaction energies calculated at the CCSDT and CCSD(T) levels of theory, in smaller basis sets (aug-cc-pV$n$Z-PP, with $n =$ D, T, Q, and aug-cc-pwCVDZ-PP). Analogously, we estimate the magnitude of noniterative and iterative quadruple excitations [$\Delta$(Q) and $\Delta$Q] using the CCSDT(Q) and CCSDTQ methods, respectively, with the aug-cc-pVDZ-PP and aug-cc-pVTZ-PP basis sets. For completeness, we also compare the PECs obtained within the coupled-cluster method with the ones calculated using the multireference configuration interaction method restricted to single and double excitations (MRCISD).


The permanent electric dipole moments and static electric dipole polarizabilities are calculated using the finite-field method at the CCSD(T)/aug-cc-pwCV5Z level of theory. The $z$ axis is chosen along the internuclear axis and oriented from a Zn or Cd atom to an alkali-metal or alkaline-earth-metal atom.

All electronic structure calculations are performed using the MOLPRO package of \textit{ab initio} programs~\cite{MOLPRO-WIREs,MOLPRO}. The full triple and quadruple contributions are computed using the MRCC code embedded in MOLPRO~\cite{MRCC}. Vibrational eigenstates are calculated numerically by employing the exact diagonalization of the nuclear motion Hamiltonian within the discrete variable representation (DVR) on the nonequidistant grid~\cite{DVR}. Atomic masses of the most abundant isotopes are assumed.

\section{Results and discussion}
\label{sec:results}
\subsection{Potential-energy curves}

\begin{figure*}[!tb]
\begin{center}
\includegraphics[width=0.95\linewidth]{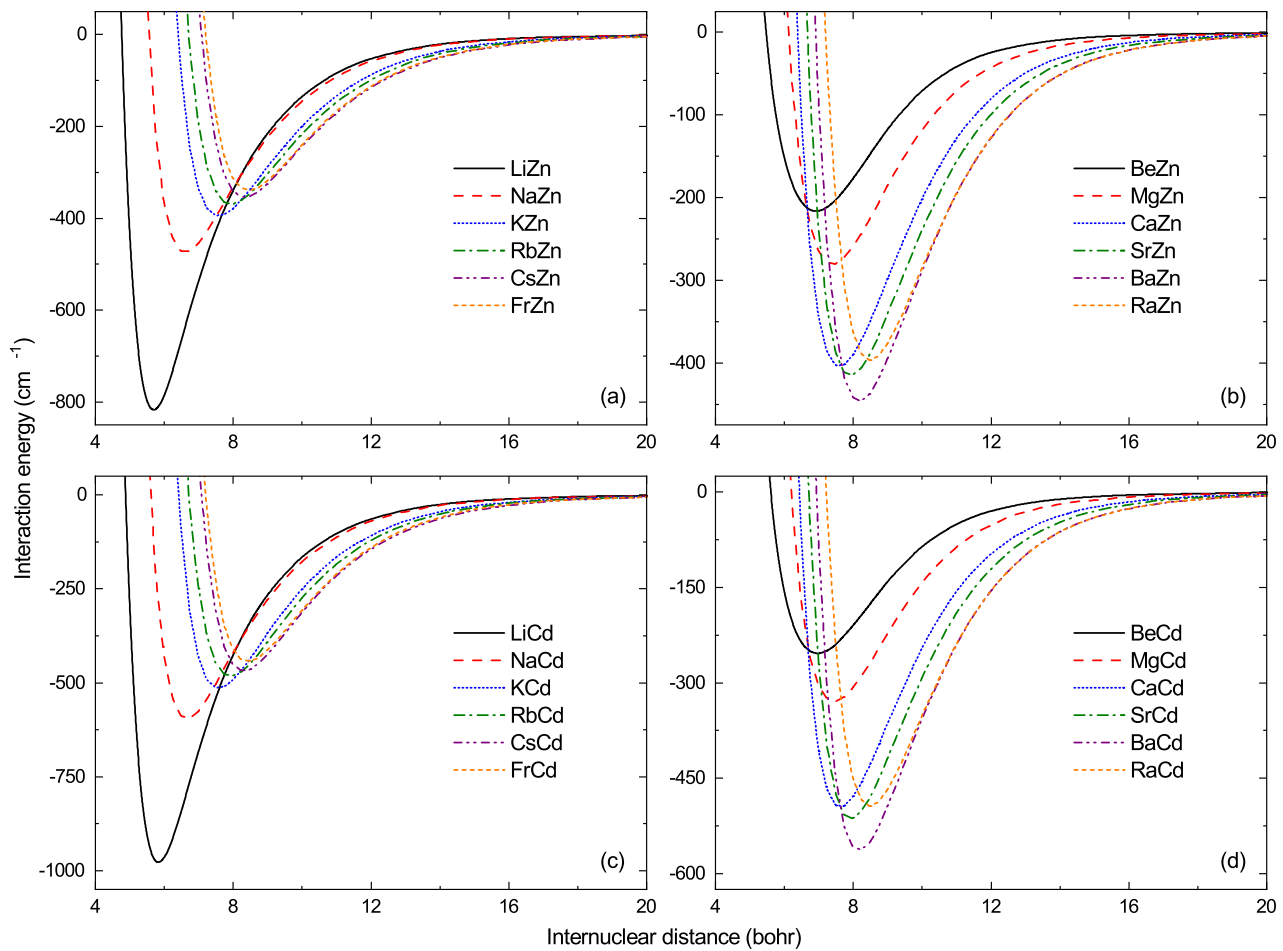}
\end{center}
\caption{Potential-energy curves of (a) the $AM$Zn molecules in the $X^2\Sigma^+$ electronic state, (b) the $AEM$Zn molecules in the $X^1\Sigma^+$ electronic state, (c) the $AM$Cd molecules in the $X^2\Sigma^+$ electronic state, and (d) the $AEM$Cd molecules in the $X^1\Sigma^+$ electronic state.}	
\label{fig:potentials}
\end{figure*}

\begin{table*}[!tb]
\caption{Spectroscopic characteristics of the $AM$Zn and $AM$Cd molecules in the $X\,^2\Sigma^+$ electronic state and $AEM$Zn and $AEM$Cd molecules in the $X\,^1\Sigma^+$ electronic state: equilibrium bond length $R_e$, well depth $D_e$, harmonic constant $\omega_e$, first anharmonicity constant $\omega_e x_e$, number of bound vibrational states $N_\nu$, rotational constant $B_e$, permanent electric dipole moment $d_e$, parallel and perpendicular components of the static electric dipole polarizability $\alpha^\parallel_e$ and $\alpha^\perp_e$, and long-range dispersion-interaction coefficient $C_6$. The results for the Zn$_2$, Cd$_2$, and ZnCd molecules are also presented.\label{tab:characteristics}}
\begin{ruledtabular}
\begin{tabular}{lccccccccccc}
Molecule & State & $R_e$ (bohr) & $D_e$ (cm$^{-1}$) & $\omega_e$ (cm$^{-1}$) & $\omega_e x_e$ (cm$^{-1}$) & $N_\nu$ & $B_e$ (cm$^{-1}$) & $d_e$ (D) & $\alpha^\parallel_e$ (a.u.) &   $\alpha^\perp_e$ (a.u.) & $C_6$ (a.u.)~\cite{QiaoJCP12} \\ \hline
	LiZn & $X\,^2\Sigma^+$ & 5.67 & 827 & 131 & 6.84 & 17 & 0.296 & 0.30 & 375 & 164 & 541 \\
	NaZn & $X\,^2\Sigma^+$ & 6.61 & 474 & 50.9& 1.51 & 24 & 0.081 & 0.20 & 315 & 173 & 597\\
	KZn & $X\,^2\Sigma^+$ & 7.57 & 395 & 34.8 & 0.35 & 28 & 0.043 & 0.18 & 457 & 291 & 837 \\
	RbZn & $X\,^2\Sigma^+$ & 7.93 & 368 & 26.6& 0.45 & 34 & 0.026 & 0.16 & 481 & 321 & 959 \\
	CsZn & $X\,^2\Sigma^+$ & 8.33 & 354 & 23.4& 0.62 & 37 & 0.020 & 0.12 & 558 & 390 & 1129 \\
	FrZn & $X\,^2\Sigma^+$ & 8.45 & 339 & 21.5& 0.41 & 39 & 0.017 & 0.11 & 467 & 331& 1056\\
	BeZn & $X\,^1\Sigma^+$ & 6.91 & 218 & 39.9& 1.88 & 11 & 0.159 &-0.03 & 101 & 69 & 270 \\
	MgZn & $X\,^1\Sigma^+$ & 7.39 & 280 & 33.2& 0.46 & 18 & 0.063 & -0.003 & 148 & 99 & 450\\
	CaZn & $X\,^1\Sigma^+$ & 7.60 & 404 & 33.4& 0.30 & 28 & 0.042 & -0.08 & 271 & 175 & 771\\
	SrZn & $X\,^1\Sigma^+$ & 7.92 & 415 & 27.4& 0.23 & 35 & 0.026 & -0.07 & 321 & 215 & 916\\
	BaZn & $X\,^1\Sigma^+$ & 8.17 & 446 & 25.7& 0.70 & 40 & 0.021 & -0.12 & 417 & 287 & 1138 \\
	RaZn & $X\,^1\Sigma^+$ & 8.49 & 396 & 22.2& 0.21 & 41 & 0.017 & -0.02 & 373 & 266 & 1044 \\
	LiCd & $X\,^2\Sigma^+$ & 5.80 & 988 & 134& 5.41 & 19 & 0.270 & 0.54 & 409 & 165 & 708 \\
	NaCd & $X\,^2\Sigma^+$ & 6.66 & 596 & 53.2& 1.19 & 28 & 0.071 & 0.40 & 350 & 175 & 783 \\
	KCd & $X\,^2\Sigma^+$ & 7.58 & 515 & 36.4& 1.50 & 34 & 0.036 & 0.44 & 511 & 288 & 1090 \\
	RbCd & $X\,^2\Sigma^+$ & 7.92 & 482 & 26.6& 0.30& 44 & 0.020 & 0.42 & 537 & 318 & 1251 \\
	CsCd & $X\,^2\Sigma^+$ & 8.32 & 467 & 22.7& 0.33 & 50 & 0.014 & 0.40 & 624 & 383 & 1470 \\
	FrCd & $X\,^2\Sigma^+$ & 8.43 & 443 & 20.1& 0.31 & 54 & 0.011 & 0.34 & 519 & 328 & 1381 \\
	BeCd & $X\,^1\Sigma^+$ & 6.95 & 255 & 42.2& 1.59 & 12 & 0.149 & -0.03 & 115 & 75 & 365 \\
	MgCd & $X\,^1\Sigma^+$ & 7.44 & 329 & 33.8& 0.48 & 21 & 0.055 & 0.02 & 164 & 105 & 605 \\
	CaCd & $X\,^1\Sigma^+$ & 7.62 & 496 & 34.5& 1.38 & 33 & 0.035 & -0.02 & 299 & 179 & 1023 \\
	SrCd & $X\,^1\Sigma^+$ & 7.93 & 513 & 26.7& 0.12 & 44 & 0.019 & 0.01 & 352 & 219 & 1212 \\
	BaCd & $X\,^1\Sigma^+$ & 8.17 & 563 & 24.7& 0.10 & 53 & 0.014 & -0.02 & 456 & 290 & 1499\\
	RaCd & $X\,^1\Sigma^+$ & 8.49 & 494 & 20.4& 0.20 & 55 & 0.011 & 0.08 & 406 & 269 & 1335 \\
	Zn$_2$ & $X\,^1\Sigma^+_g$ & 7.23 & 231 & 23.4& 0.62& 22 & 0.036 & 0 & 97 & 70 & 359 \\
	ZnCd & $X\,^1\Sigma^+$ & 7.28 & 275 & 22.6 & 0.53 & 27 & 0.028 & 0.01 & 110 & 76 & 495 \\
	Cd$_2$ & $X\,^1\Sigma^+_g$ & 7.32 & 330 & 21.1 & 0.24 & 35 & 0.020 & 0 & 124 & 83 & 686 \\
\end{tabular} 
\end{ruledtabular}
\end{table*}

We consider interactions between a zinc or cadmium atom and an alkali-metal $AM$ ($AM$ = Li, Na, K, Rb, Cs, Fr) or alkaline-earth-metal $AEM$ ($AEM$ = Be, Mg, Ca, Sr, Ba, Ra) atom in their electronic ground states. The ground-state Zn and Cd atoms, as well as alkaline-earth-metal atoms, are described with the $^1 S_0$ electronic term, while the alkali-metal atoms are described with the $^2 S_{1/2}$ electronic term. This yields the $^2\Sigma^+$ molecular electronic states for the ground-state open-shell molecules composed of a Zn or Cd atom and an alkali-metal atom and the $^1 \Sigma^+$ molecular electronic states for the ground-state closed-shell molecules composed of a Zn or Cd atom and an alkaline-earth-metal atom. 

For the above molecules, we compute the potential-energy curves and provide spectroscopic characteristics: the equilibrium bond lengths $R_e$, potential-well depths $D_e$, harmonic constants $\omega_e$, first anharmonicity constants $\omega_e x_e$, numbers of bound vibrational states $N_\nu$, and rotational constants $B_e$. We also report the permanent electric dipole moments $d_e$ and parallel and perpendicular components of the static electric dipole polarizabilities, $\alpha^\parallel_e$ and $\alpha^\perp_e$, respectively, at equilibrium distances. The computed curves are presented in Fig.~\ref{fig:potentials}, and obtained characteristics are collected in Table~\ref{tab:characteristics}. We estimate the number of bound vibrational states $N_\nu$ with the use of the DVR method, in which we employ the computed PECs, describing the short-range part of the interaction, smoothly connected with the long-range part of the interaction, $-C_6/R^6$, where the dispersion-interaction coefficients $C_6$ are taken from Ref.~\cite{QiaoJCP12} and presented in Table~\ref{tab:characteristics} for completeness. We also provide the results for the ground-state homonuclear Zn$_2$ and Cd$_2$ dimers described with the $^1 \Sigma_g^+$ molecular term and the ZnCd molecule in the $^1 \Sigma^+$ electronic ground state.

Figure~\ref{fig:potentials} presents the potential-energy curves of the $AM$Zn and $AM$Cd molecules in the $X\,^2\Sigma^+$ electronic ground state, and $AEM$Zn and $AEM$Cd molecules in the $X\,^1\Sigma^+$ electronic ground state calculated at the CCSD(T)+$\Delta$T level of theory. All PECs exhibit a smooth behavior with well-defined minima. For the $AM$Zn and $AM$Cd molecules, the well depths systematically decrease with the increasing atomic number of the alkali-metal atoms (a deviation from the trend is observed for radium-containing molecules), while for the $AEM$Zn and $AEM$Cd molecules, the well depths systematically increase with the increasing atomic number of the alkaline-earth-metal atoms. The opposite trends can be explained by different characters of bonding within molecules containing alkali-metal and alkaline-earth atoms: the open-shell $AM$Zn and $AM$Cd molecules are bound chemically (with a bond order of $\frac{1}{2}$), while the closed-shell $AEM$Zn and $AEM$Cd molecules are bound solely by the dispersion forces. All considered molecules are of van der Waals character, with moderate equilibrium distances and well depths not exceeding 1000$\,$cm$^{-1}$. We also notice that molecules containing cadmium are more strongly bound than molecules containing zinc due to the larger polarizability of the cadmium atom. 

The well depths of the $AM$Zn molecules in the ground $X\,^2\Sigma^+$ electronic state range from 827$\,$cm$^{-1}$ for LiZn to 339$\,$cm$^{-1}$ for FrZn, systematically decreasing with the atomic number of the alkali-metal atom $AM$. The equilibrium distances range from 5.67$\,$bohrs for LiZn to 8.45$\,$bohrs for FrZn, systematically increasing with the atomic number of $AM$. We observe the same trend for the $AM$Cd molecules in the ground $X\,^2\Sigma^+$ electronic state, whose well depths range from 988$\,$cm$^{-1}$ for LiCd to 433$\,$cm$^{-1}$ for FrCd, and equilibrium distances increase from 5.80$\,$bohrs for LiCd to 8.43$\,$bohrs for FrCd. The computed number of vibrational levels increases with the reduced mass of the molecule, from 17 and 19 for LiZn and LiCd to 39 and 54 for FrZn and FrCd, respectively.

For the $AEM$Zn molecules in the ground $X\,^1\Sigma^+$ electronic state, the well depth systematically increases from 218$\,$cm$^{-1}$ for BeZn to 446$\,$cm$^{-1}$ for BaZn and slightly drops to 396$\,$cm$^{-1}$ for RaZn. $AEM$Cd molecules in the ground $X\,^1\Sigma^+$ electronic state are characterized by well depths which also systematically increase with the atomic number of $AEM$, ranging from 255$\,$cm$^{-1}$ for BeCd to 563$\,$cm$^{-1}$ for BaCd and 494$\,$cm$^{-1}$ for RaCd. The equilibrium distance systematically increases from 6.91$\,$bohrs for BeZn to 8.49$\,$bohrs for RaZn and from 6.95$\,$bohrs for BeCd to 8.49$\,$bohrs for RaCd. The estimated number of vibrational levels amounts to 11 and 12 for BeZn and BeCd and increases with the reduced mass of the molecule up to 41 and 55 for RaZn and RaCd, respectively.

The observed trends in the studied molecules are similar to those reported for analogous alkali-metal--alkaline-earth-metal and alkaline-earth-metal molecules~\cite{GueroutPRA10,HeavenCPL11,PototschnigPCCP16}. However, the potential-well depths are smaller, and equilibrium distances are larger in the present case due to smaller polarizabilities of the Zn and Cd atoms than those of alkaline-earth-metal atoms.

\begin{figure}[tb!]
\begin{center}
\includegraphics[width=1\linewidth]{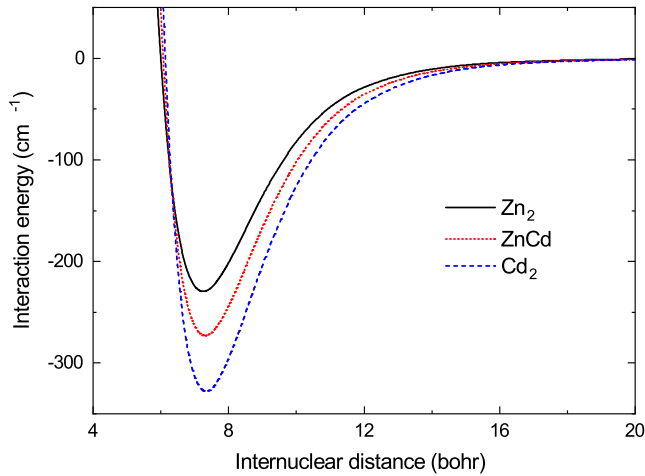}
\end{center}
\caption{Potential-energy curves of the Zn$_2$ and Cd$_2$ molecules in the $X\,^1\Sigma^+_g$ electronic state and the ZnCd molecule in the $X\,^1\Sigma^+$ electronic state.}
\label{fig:potentials_ZnCd}
\end{figure}

For completeness, we also provide results for the homonuclear Zn$_2$ and Cd$_2$ and heteronuclear ZnCd molecules in their ground $X\,^1\Sigma^+_g$ and $X\,^1\Sigma^+$ electronic states, respectively. The calculated spectroscopic characteristics are collected in Table~\ref{tab:characteristics}. Figure~\ref{fig:potentials_ZnCd} presents the PECs, which were calculated at the CCSD(T)+$\Delta$T+$\Delta$(Q) level of theory. The potential-well depths amount to 231, 275, and 330$\,$cm$^{-1}$ for Zn$_2$, ZnCd, and Cd$_2$, respectively, and the respective equilibrium distances are 7.23, 7.28, and$\,$7.32 bohrs. The estimated number of vibrational levels is equal to 22 for Zn$_2$, 27 for ZnCd, and 35 for Cd$_2$. The values of the well depths $D_e$ and harmonic constants $\omega_e$ obtained for homonuclear dimers are in good agreement with the results of previous theoretical calculations and spectroscopic measurements, which are compared in Table~\ref{tab:Zn2Cd2}. Like previous theoretical works, we observe discrepancies between the calculated equilibrium distances and their experimental values, especially for Zn$_2$.

\begin{table}[h!]
\caption{Spectroscopic constants of the Zn$_2$ and Cd$_2$ molecules in the $X\,^1\Sigma^+_g$ electronic state: Comparison with previous studies.}
\label{tab:Zn2Cd2}
\begin{ruledtabular}
\begin{tabular}{llccc}
Molecule & Source & $R_e\,$(bohr) & $D_e\,$(cm$^{-1}$) & $\omega_e$ (cm$^{-1}$)\\
\hline 
   Zn$_2$ & This work & 7.23 & 231 & 23.4  \\ 
          & Theory~\cite{PetersonTCA05} & 7.27 & 226 &  23.9 \\ 
	  & Theory~\cite{PahlTCA11} & 7.23 & 226 & 24.0  \\ 
	  & Theory~\cite{WeiJCP13} & 7.32 & 242 & 25.65 \\ 
	  & Experiment~\cite{StrojeckiCP06} & 7.92 & 242 & 25.9  \\ 
   Cd$_2$ & This work & 7.32 & 330 & 21.1 \\ 
          & Theory~\cite{PetersonTCA05} & 7.36 & 325 & 20.2  \\ 
	  & Theory~\cite{PahlTCA11} & 7.32 & 319 &  21.3 \\ 
	  & Theory~\cite{WeiJCP13} & 7.75 & 328 & 21.5  \\ 
	  & Experiment~\cite{StrojeckiCPL10} & 7.14$\pm$0.06 & 328$\pm$3 & 21.4$\pm$0.2  \\ 
\end{tabular} 
\end{ruledtabular}
\end{table}

\subsection{Permanent electric dipole moments and static electric dipole polarizabilities}

\begin{figure*}[!tb]
\begin{center}
\includegraphics[width=0.95\linewidth]{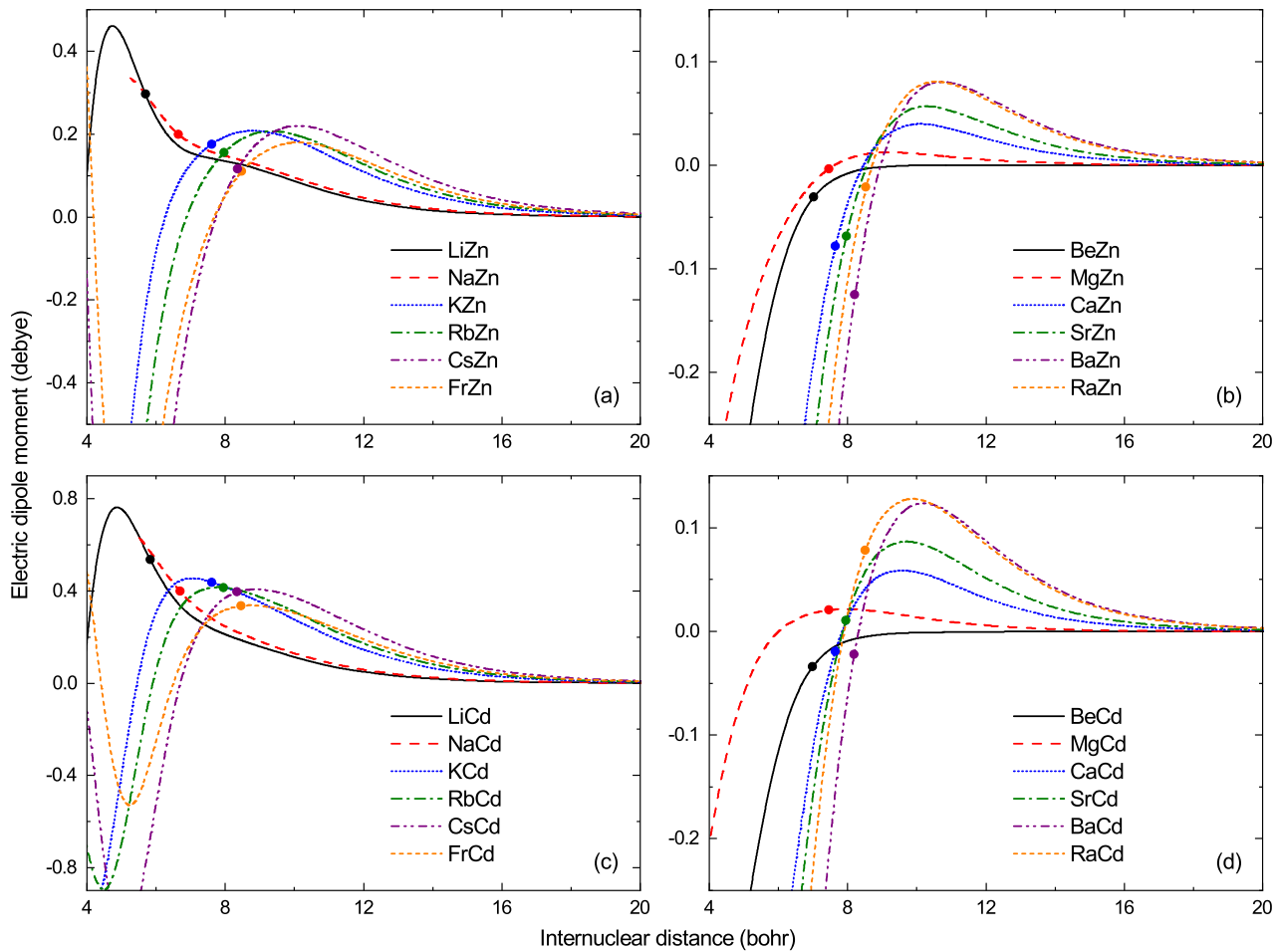}
\end{center}
\caption{Permanent electric dipole moments of (a) the $AM$Zn molecules in the $X^2\Sigma^+$ electronic state, (b) the $AEM$Zn molecules in the $X^1\Sigma^+$ electronic state, (c) the $AM$Cd molecules in the $X^2\Sigma^+$ electronic state, and (d) the $AEM$Cd molecules in the $X^1\Sigma^+$ electronic state. The points mark the permanent electric dipole moments at equilibrium distances.}
\label{fig:dipoles}
\end{figure*}

The permanent electric dipole moments of the $AM$Zn and $AM$Cd molecules in the $X\,^2\Sigma^+$ electronic state and the $AEM$Zn and $AEM$Cd molecules in the $X\,^1\Sigma^+$ electronic states as functions of the internuclear distance are presented in Fig.~\ref{fig:dipoles}. The values of permanent electric dipole moments at equilibrium distances are collected in Table~\ref{tab:characteristics}. They govern the strength of the intermolecular dipolar interaction and coupling with an external static electric field. 

The $AM$Zn and $AM$Cd molecules in the $X\,^2\Sigma^+$ electronic state have small permanent EDMs, not exceeding 0.54 debye at equilibrium distances. The values of the EDMs at equilibrium distances systematically decrease with the increasing atomic number of the alkali-metal atom, ranging from  0.30 debye for LiZn to 0.11 debye for FrZn and from 0.54 for LiCd to 0.34 for FrCd.

The permanent EDMs of the $AEM$Zn and $AEM$Cd molecules in the $X\,^1\Sigma^+$ electronic state take even smaller values, not exceeding 0.12 debye at equilibrium distances. The equilibrium-distance EDMs take values ranging from 0.003 debye for MgZn to 0.12 debye for BaZn and from 0.01 for SrCd to 0.08 for RaCd, with no distinct systematics of atomic number dependence. In contrast to the $AM$Zn molecules, the permanent electric dipole moments of the $AEM$Zn molecules point from the alkaline-earth-metal atom to the zinc atom. For BeCd, CaCd, and BaCd molecules, the dipoles are oriented from the alkaline-earth-metal atom to the cadmium atom, while for MgCd, SrCd, and RaCd, the dipoles are oriented from the cadmium atom to the alkaline-earth-metal atom. 

The permanent electric dipole moments of all the studied molecules take very small values despite relatively large electronegativity differences between involved atoms. The electronegativity by the Pauling scale of the Zn (1.65) and Cd (1.69) atoms is almost two times larger than that of the alkali-metal (0.79-0.98) and alkaline-earth-metal (0.89-1.57) atoms~\cite{Pauling1960}. For such large differences, significant permanent EDMs, larger than those for analogous alkali-metal--alkaline-earth-metal and alkaline-earth-metal molecules~\cite{GueroutPRA10,HeavenCPL11,PototschnigPCCP16}, could be expected (similar to what was recently reported for molecules containing Cu and Ag atoms~\cite{SmialkowskiPRA21}). Unfortunately, the present values are significantly smaller, partially as a result of weak interatomic interactions and van der Waals nature of the studied molecules and partially because around minima the permanent EDMs cross zero and change sign. At large distances, the permanent EDMs present an expected systematic dependence on the electronegativity differences. 

We also calculate the parallel, $\alpha^\parallel_e$, and perpendicular, $\alpha^\perp_e$, components of the static electric dipole polarizability tensor, which play an important role in the evaluation of intermolecular interactions and coupling of molecular rovibrational dynamics with a laser field. The values of the parallel and perpendicular components of static electric dipole polarizabilities at equilibrium distances are collected in Table~\ref{tab:characteristics}.

\subsection{Convergence and accuracy analysis}

In order to investigate the uncertainty of the present molecular electronic structure calculations, we first examine whether the employed \textit{ab initio} methods describe the atomic properties accurately. To this end, we employ the CCSD(T) method to calculate the atomic static electric dipole polarizabilities and ionization potentials of the Zn and Cd atoms using the aug-cc-pwCV5Z-PP basis sets. The obtained atomic polarizabilities of the zinc and cadmium atoms amount to 37.7~and 45.8$\,$a.u.~and agree well with the recommended combined experimental-theoretical values of 38.7$\pm$0.3~and 46$\pm$2$\,$a.u.~\cite{SchwerdtfegerMP19}, respectively. The calculated ionization potentials of 75848$\,$cm$^{-1}$ for Zn and 72526$\,$cm$^{-1}$ for Cd are also in good agreement with respective experimental values of 75769$\,$cm$^{-1}$~\cite{SugarJCP95} and 72540$\,$cm$^{-1}$~\cite{ShenstoneOSA49}. The atomic polarizabilities and ionization potentials of the alkali-metal and alkaline-earth-metal atoms obtained with the used theoretical methods are also in good agreement with experimental data, as confirmed in Ref.~\cite{SmialkowskiPRA21}.

\begin{figure}[!tb]
\begin{center}
\includegraphics[width=1\linewidth]{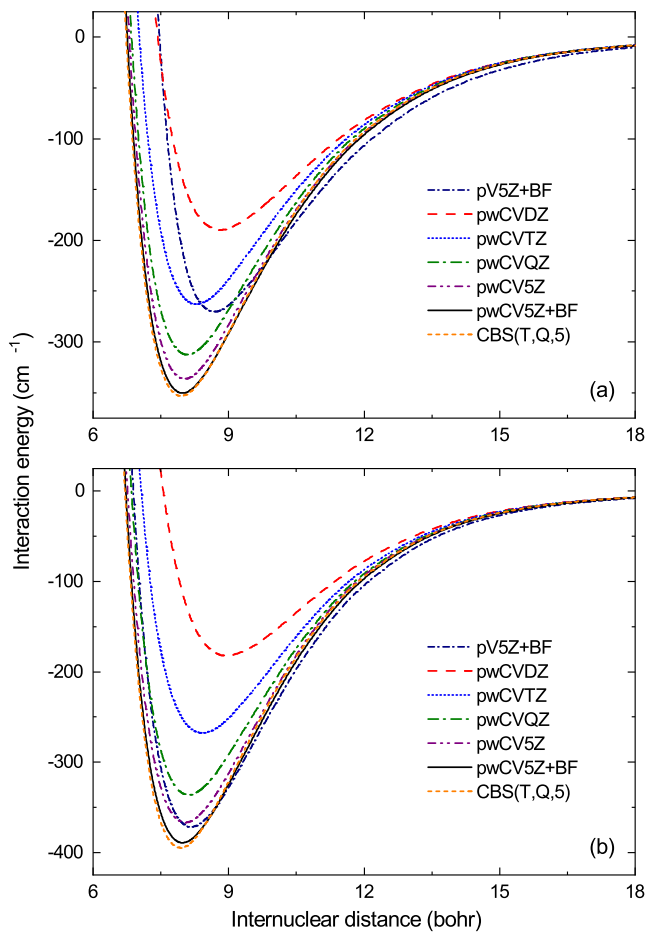}
\end{center}
\caption{Potential-energy curves of (a)~the RbZn molecule in the $X\,^2\Sigma^+$ electronic state and (b)~the SrZn molecule in the $X\,^1\Sigma^+$ electronic state, computed with the CCSD(T) method, using different-sized Gaussian basis sets. CBS limit energy for the aug-cc-pwCV$n$Z basis sets is also presented.}
\label{fig:conv_basis}
\end{figure}

Next, we analyze the convergence of the interatomic interaction energy with the size of the employed basis sets. Figure~\ref{fig:conv_basis} presents the potential-energy curves for the representative RbZn and SrZn molecules in the $X\,^2\Sigma^+$ and $X\,^1\Sigma^+$ electronic states, respectively, obtained with different basis sets. An inspection of Fig.~\ref{fig:conv_basis} allows us to conclude that the inclusion of inner-shell electron correlation is crucial for an accurate description of the interatomic interactions and the core-core and core-valence contributions are significant, especially for the $AM$Zn and $AM$Cd molecules. The addition of a bond function to the aug-cc-pwCV5Z-PP basis set allows describing the complete-basis-set-limit energy accurately.

\begin{figure}[!tb]
\begin{center}
\includegraphics[width=1\linewidth]{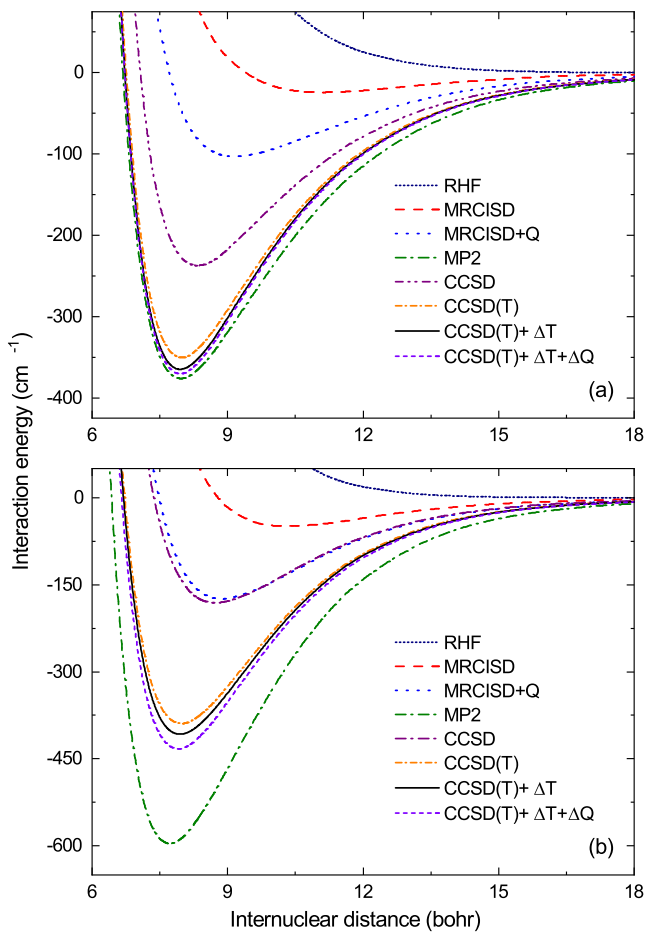}
\end{center}
\caption{Potential energy curves of (a)~the RbZn molecule in the $X\,^2\Sigma^+$ electronic state and (b)~the SrZn molecule in the $X\,^1\Sigma^+$ electronic state computed at different levels of theory: RHF, MRCISD, MRCISD+Q, MP2, CCSD, CCSD(T), CCSD(T)+$\Delta$T, and CCSD(T)+$\Delta$T+$\Delta$Q. See the text for details.}
\label{fig:conv_method}
\end{figure}

Finally, we analyze the convergence of the interatomic interaction energy with the quality of employed wave-function representation. Figure~\ref{fig:conv_method} presents the potential-energy curves for the representative RbZn and SrZn molecules in the $X\,^2\Sigma^+$ and $X\,^1\Sigma^+$ electronic states, respectively, calculated at different levels of theory: spin-restricted Hartree-Fock RHF, MRCISD, MRCISD+Q, second-order M\o{}ller-Plesset perturbation theory MP2, CCSD, CCSD(T), CCSD(T)+$\Delta$T, and CCSD(T)+$\Delta$T+$\Delta$Q. The aug-cc-pwCV5Z-PP+BF basis set is used in the CCSD(T) calculations, aug-cc-pVTZ-PP is used in the calculation of the full triple correction $\Delta$T, and aug-cc-pVDZ-PP is used in the calculation of the full quadruple correction $\Delta$Q.

In the coupled-cluster calculations, the inclusion of higher-order excitations significantly improves the description of the interaction energies. The obtained potential-well depths for the RbZn and SrZn molecules are equal to, respectively, 352 and 391$\,$cm$^{-1}$ at the CCSD(T) level, 368 and 415$\,$cm$^{-1}$ at the CCSD(T)+$\Delta$T level, and 372 and 435$\,$cm$^{-1}$ at the CCSD(T)+$\Delta$T+$\Delta$Q level. Hence, the full triple correction increases the well depth by about 4\% for RbZn and about 6\% for SrZn, while the full quadruple correction introduces a further 1\% well-depth increase for RbZn and about 5\% for SrZn. Moreover, for the Zn$_2$, ZnCd, and Cd$_2$ dimers, the perturbative quadruple correction $\Delta$(Q) (calculated with the use of the aug-cc-pVTZ-PP basis set) increases the potential-well depths obtained at the CCSD(T)+$\Delta$T level by 5\% to 7\%. 

Therefore, an accurate description of the interatomic interactions between closed-shell transition-metal atoms, like zinc and cadmium, and alkali-metal or alkaline-earth-metal atoms has to take into consideration higher-order electron correlation, in contrast to alkali-metal dimers with two valence electrons~\cite{GronowskiPRA20}. The triple-excitation contribution is significant for both open-shell $AM$Zn and $AM$Cd and closed-shell $AEM$Zn and $AEM$Cd molecules, while the quadruple-excitations contribution is particularly large for closed-shell $AEM$Zn and $AEM$Cd molecules. It is in agreement with the fact that the CCSDT method for the $AM$Zn and $AM$Cd molecules with three valence electrons already provides the description of valence electrons at the full configuration interaction level, while the $AEM$Zn and $AEM$Cd molecules with four valence electrons require the CCSDTQ method for the same.

We also find that the values of the full-iterative triple and full-iterative quadruple corrections systematically decrease with the increasing size of the basis sets, leading us to conclude that the calculated corrections may be slightly overestimated. Overall, we estimate that the uncertainty of our calculations should be at most 5\% for molecules containing alkali-metal atoms and slightly more for molecules containing alkaline-earth atoms.

In the MRCISD calculations, the full-valence active space is used. The well depths of the potential-energy curves obtained at the MRCISD level amount to 22.8$\,$cm$^{-1}$ for RbZn and 51.5$\,$cm$^{-1}$ for SrZn. The addition of the Davidson correction, MRCISD+Q, yields deeper PECs, with the well depths of 103 and 175$\,$cm$^{-1}$, respectively, for RbZn and SrZn, yet the results are still not comparable to those obtained with the coupled-cluster method. We also calculate the PECs within the MRCISD method with $4p$ Zn orbitals and $5p$ Rb/Sr orbitals included in the active space; however, this approach leads to almost identical results. Interestingly, those PECs do not differ significantly from the ones obtained at the CISD level, meaning that the ground electronic states of RbZn and SrZn are well described by a single reference, and higher-order excitations need to be taken into account to reproduce the coupled-cluster results. We also see that the energies obtained within the second-order M{\o}ller-Plesset perturbation theory (MP2) are overestimated, particularly for the SrZn molecule.

\subsection{Chemical reactions}

We use the calculated potential-well depths $D_e$ to assess the stability of the investigated molecules against atom-exchange chemical reactions~\cite{ZuchowskiPRA10,TomzaPRA13,SmialkowskiPRA20}. A ground-state heteronuclear molecule $AB$ can undergo an atom-exchange chemical reaction,
\begin{equation}
AB + AB \rightarrow A_2 + B_2,
\end{equation}
provided that the sum of the dissociation energies $D_0$ of the products, $A_2$ and  $B_2$, is larger than or equal to the sum of the dissociation energies of the reactants, $2 AB$, 
\begin{equation}
D_0(A_2) + D_0(B_2) \ge 2 D_0(AB).
\end{equation}
The dissociation energy $D_0$ is related to the potential-well depth $D_e$, $D_0 \approx D_e - \frac{1}{2} \omega_e$. 

All the studied $AM$Zn, $AM$Cd, $AEM$Zn, and $AEM$Cd molecules in the ground rovibrational levels of their ground electronic states are chemically unstable and atom-exchange reactions are energetically possible,
\begin{equation}
\begin{split}
2 AM\mathrm{Zn}(X^2\Sigma^+) &\to \mathrm{Zn}_2 (X^1\Sigma^+_g) + AM_2 (X^1\Sigma^+_g) \,,\\
2 AM\mathrm{Cd}(X^2\Sigma^+) &\to \mathrm{Cd}_2 (X^1\Sigma^+_g) + AM_2 (X^1\Sigma^+_g) \,,\\
2 AEM\mathrm{Zn}(X^1\Sigma^+) &\to \mathrm{Zn}_2 (X^1\Sigma^+) + AEM_2 (X^1\Sigma^+_g) \,,\\
2 AEM\mathrm{Cd}(X^1\Sigma^+) &\to \mathrm{Cd}_2 (X^1\Sigma^+_g) + AEM_2 (X^1\Sigma^+_g) \,,
\end{split}
\end{equation}
because the well depths of alkali-metal $AM_2$ and alkaline-earth-metal $AEM_2$ dimers~\cite{ZuchowskiPRA10} are significantly larger than those of the present molecules. 

The atom-exchange chemical reactions for the $AM$Zn and $AM$Cd molecules could potentially be suppressed by restricting the collision dynamic to the high-spin potential-energy surfaces by fully spin polarizing the molecules in an external magnetic field~\cite{TomzaPRA13}. Unfortunately, only channels leading to alkali-metal dimers in the $X^3\Sigma^+_u$ electronic state are closed, while ones leading to Cd$_2$ and Zn$_2$ in the $X^3\Sigma^+_u$ electronic state are open,
\begin{equation}
\begin{split}
2 AM\mathrm{Zn}(X^2\Sigma^+) & \not\to \mathrm{Zn}_2 (X^1\Sigma^+_g) + AM_2 (a^3\Sigma^+_u) \,,\\
2 AM\mathrm{Cd}(X^2\Sigma^+) & \not\to \mathrm{Cd}_2 (X^1\Sigma^+_g) + AM_2 (a^3\Sigma^+_u) \,.\\
2 AM\mathrm{Zn}(X^2\Sigma^+) & \to \mathrm{Zn}_2 (a^3\Sigma^+_u) + AM_2 (X^1\Sigma^+_g) \,,\\
2 AM\mathrm{Cd}(X^2\Sigma^+) & \to \mathrm{Cd}_2 (a^3\Sigma^+_u) + AM_2 (X^1\Sigma^+_g) \,.\\
\end{split}
\end{equation}

Finally, most likely, the trimer formation reactions are another path of chemical losses for all considered molecules~\cite{ZuchowskiPRA10,TomzaPRA13,SmialkowskiPRA20},
\begin{equation}
AB + AB \to A_2B + B,
\end{equation}
but their detailed study is out of the scope of this work.

\section{Summary and conclusions}
\label{sec:summary}

Motivated by the recent progress in laser cooling and trapping of cadmium atoms~\cite{XuPRA04,BrickmanPRA07,KanedaOL16,YamaguchiPRA19} and experimental realizations of ultracold mixtures of closed-shell and open-shell atoms~\cite{NemitzPRA09,BarbeNP18,GreenPRX20,WilsonPRA21}, in this paper, we brought attention to diatomic molecules composed of a closed-shell zinc or cadmium atom interacting with an alkali-metal (Li, Na, K, Rb, Cs, Fr) or alkaline-earth-metal (Be, Mg, Ca, Sr, Ba, Ra) atom. Such molecules are potential candidates for ultracold quantum physics and chemistry experiments, ranging from controlled chemical reactions to precision measurements. To this end, we have carried out state-of-the-art \textit{ab initio} calculations of the potential energy curves, permanent electric dipole moments, and spectroscopic constants for the molecules in their electronic ground states. We have used the \textit{ab initio} electronic structure coupled-cluster method with single, double, and triple excitations combined with large Gaussian basis sets and small-core relativistic energy-consistent pseudopotentials for heavier atoms.  We have predicted that the studied molecules in the ground electronic state are weakly bound van der Waals complexes. We have also found that they have rather small permanent electric dipole moments, despite Zn and Cd atoms having electronegativity significantly larger than that of alkali-metal and alkaline-earth-metal atoms. Finally, we have concluded that they are chemically reactive, and for applications other than studies of ultracold chemical reactions, they should be segregated in an optical lattice, or shielding strategies should be employed~\cite{Anderegg2021}.
 
Full potential energy curves and permanent electric dipole moments as a function of interatomic distance in a numerical form are collected in the Supplemental Material~\footnote{See Supplemental Material at http://link.aps.org/supplemental/XXXX for the calculated potential energy curves and permanent electric dipole moments in a numerical form.}



\begin{acknowledgments}
We acknowledge the financial support from the National Science Centre of Poland (Grant No. 2016/23/B/ST4/03231) and the Foundation for Polish Science within the First Team program co-financed by the European Union under the European Regional Development Fund. The computational part of this research was partially supported by the PL-Grid Infrastructure.
\end{acknowledgments}

\bibliography{ZnCd}

\end{document}